\documentclass[preprint,5p]{elsarticle}

\usepackage{amssymb}
\usepackage{float} 
\usepackage{caption}
\usepackage{subcaption}
\usepackage{graphicx, booktabs}
\usepackage{graphics}
\usepackage{amsmath}
\usepackage{pdfpages}
\usepackage{mathtools}
\usepackage{setspace}
\usepackage{rotating}    
\usepackage{chngcntr}
\usepackage{etoolbox}
\usepackage{multirow}
\usepackage{makecell}
 \usepackage{color}
 \usepackage{vcell}
 \usepackage{siunitx}
\usepackage{algorithm}
\usepackage{algpseudocode}
\usepackage{multicol}
\usepackage{soul}

\usepackage[acronym]{glossaries-extra}
\setabbreviationstyle[acronym]{long-short}
\newacronym{DM}{DM}{Davey-MacKay}

\journal{Digital Signal Processing}

\begin{document}

\begin{frontmatter}



\title{Watermark-Based Code Construction for Finite-State Markov Channel with Synchronisation Errors}


\author[inst1]{Shamin Achari}
% \ead{Shamin.Achari@wits.ac.za}
%\cortext[cor1]{Corresponding Author: Shamin Achari}

\affiliation[inst1]{organization={School of Electrical and Information Engineering},%Department and Organization
            addressline={University of the Witwatersrand}, 
            city={Johannesburg},
            postcode={2000}, 
            state={Gauteng},
            country={South Africa}}

\author[inst1]{Ling Cheng\corref{cor1}}
\ead{Ling.Cheng@wits.ac.za}
\cortext[cor1]{Corresponding Author: Ling Cheng}

\begin{abstract}
With advancements in telecommunications, data transmission over increasingly harsher channels that produce synchronisation errors is inevitable. Coding schemes for such channels are available through techniques such as the Davey-MacKay watermark coding; however, this is limited to memoryless channel estimates. Memory must be accounted for to ensure a realistic channel approximation - similar to a Finite State Markov Chain or Fritchman Model. A novel code construction and decoder are developed to correct synchronisation errors while considering the channel's correlated memory effects by incorporating ideas from the watermark scheme and memory modelling. Simulation results show that the proposed code construction and decoder rival the first and second-order Davey-MacKay type watermark decoder and even perform slightly better when the inner-channel capacity is higher than 0.9. The proposed system and decoder may prove helpful in fields such as free-space optics and possibly molecular communication, where harsh channels are used for communication. 
\end{abstract}


\end{frontmatter}
\let\thefootnote\relax\footnotetext{\tiny{This work has been submitted to the Elsevier DSP for possible publication. Copyright may be transferred without notice, after which this version may no longer be accessible}}
\section{Introduction}
As communication systems and technology advance, we inevitably begin to transmit data over increasingly harsher channels, and the need for digital signal processing techniques to improve the reliability of communications, as always, plays a vital role. These harsh channels naturally produce insertions and deletions (synchronisation errors) in addition to the standard substitution error. This impact is already witnessed in prevailing real-world communication systems that use visible light or free-space optics where synchronisation between the receiver and transmitter is not easily maintained and as a result, these systems suffer from synchronisation errors \cite{achari2021selfsynchronising,achari2021symbol}. The concern of synchronisation and its related errors are even found in unconventional fields such as DNA sequencing \cite{kracht2014using,kracht2015insertion} to even more intricate specialisations such as molecular communications \cite{gursoy2021towards,gomez2022age}. The latter has gained popularity in recent years as a potential for data communication in nano and micro systems, where open problems in the field contemplate the effects of synchronisation and memory \cite{gursoy2021towards,gomez2022age}.
Davey and MacKay developed a watermark coding scheme in \cite{davey2000error,mackay} to deal with such insertion, deletion and substitution (IDS) errors. While this scheme proves useful in many fields, the main shortfall is that the channel used is a memoryless approximation,  the \gls{DM} channel model, and many practical scenarios are hardly ever memoryless or independent and identically distributed (IID) in nature \cite{kanal1978models}. A recent paper by Achari et al. describes the modelling of an IDS channel that incorporates the effects of memory \cite{achari2021symbol}. Here the channel is based on a Finite-State Markov Channel (FSMC), which allows the system to transition between various states (transmission, substitution, deletion and insertion) based on different probabilities and the given current state. This provides a valuable method to model correlated synchronisation channels and also provides a method to simulate various scenarios without the need for actual transceivers. Again, while the model and method presented are helpful, it is not without drawbacks. As the model is a general method of simulating such channels, it lacks any steadfast rules and limitations, which causes challenges in producing an effective encoding/decoding algorithm to protect against and correct communication errors.

In this paper, the IDS memory model presented by Achari et al. is adapted and integrated with the DM watermark scheme to create a more encompassing channel and code construction that is more indicative of realistic communication channels while still being simple enough to provide an effective error-correction and resynchronisation scheme. The main contributions of this paper are threefold. Firstly, the new proposed channel model, which combines ideas from the DM and FSMC models, is presented. Secondly, the code construction and decoding, which are based on the watermark codes, are described for the proposed channel. Lastly, we provide the details of extending the first-order decoding for the DM watermark scheme to a second-order system. While the last contribution has been alluded to in \cite{davey2000error,mackay}, to the authors' knowledge, no explicit formulation is found in the literature. This second-order IID decoding will be used as one of the comparative benchmarks to test the proposed model and decoding algorithm against.

The rest of the paper is structured as follows. The DM channel and watermark code construction and the IDS FSMC are further detailed in Section \ref{sec:background}. This is followed by Section \ref{sec:Proposedsystem} where the novel adapted channel model is described along with the new code construction. Section \ref{sec:Simulations} then discusses the evaluation metrics and compares the simulations' various tests and results with relevant analysis. Conclusions are finally drawn in Section \ref{sec:Conclusion}.

\section{Background and Literature Review}
\label{sec:background}

\subsection{Synchronisation Channels and Coding}
Synchronisation channels remain a relatively unstudied and open topic in information theory and are generally more complex to analyse than their traditional substitution channel counterparts \cite{cheraghchi2020overview}. Mitzenmacher provides an early survey of these synchronisation channels in \cite{mitzenmacher2009survey} and appropriately draws comparisons in the analysis of synchronisation channels to traditional channels by comparing them to constructs of Levenshtein Distance over the more straightforward Hamming distance. With recent advancements in DNA-type storage, a recent surge in research regarding insertion and deletion channels has since been prompted. More recently, \cite{cheraghchi2020overview} and \cite{haeupler2021synchronization} provide detailed surveys of the progress made in the investigation of synchronisation channels. Cheraghchi et al., in particular, survey work regarding the capacity of such channels \cite{cheraghchi2020overview}. 

\subsection{Davey-Mackay Synchronisation Channel and Watermark Code}
The \gls{DM} watermark code provides a way to ensure reliable communication across a channel that produces IDS errors. A detailed description of the code construction can be found in \cite{davey2000error,mackay} and readers are encouraged to survey these texts for an in-depth explanation and analysis. What follows is the basic overview of watermark codes in order to distinguish the novelty of the proposed channel and code construction.

In the \gls{DM} construction a message string, \textbf{m}, consisting of $q$-ary symbols, are encoded using LDPC, which produces a corresponding binary sequence, \textbf{d}. This sequence \textbf{d} is then processed through a sparsifier which seeks to create a sparse sequence using a codebook or lookup table entry for each input symbol (group of $q$ bits). The mean density of the sparse vectors is denoted by $f$, and the output of the sparsifer is the sparse binary string, \textbf{s}. Here \textbf{s} is defined as sparse if the hamming weight of the sequence divided by the length is less than $0.5$. This sparse sequence is then modulo-two added with a watermark sequence, \textbf{w}, to produce \textbf{t}, which will be the sequence transmitted over the channel. The watermark sequence is one of the main components that allow the system to regain synchronisation during the decoding process. It can essentially be seen as a timing sequence as it is known to both the receiver and transmitter. The watermark sequence is generally run-length limited or random. \cite{nguyen2013watermark} Provides a method to create the watermark sequence based on a proposed probability metric. For this paper, all watermark sequences are considered random. The sequence \textbf{t} is then queued to be sent over the DM IDS channel.

\begin{figure}[th]
\centering
\includegraphics[width=0.8\columnwidth]{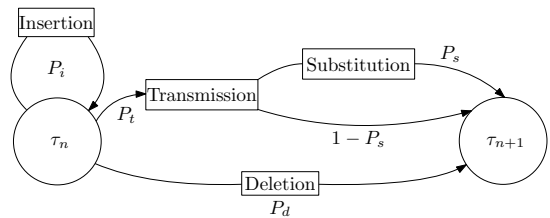}
\caption[Davey-Mackay synchronisation channel model]{Davey-Mackay synchronisation channel model}
\label{fig:DMChannel}
\end{figure}

Bits queued for transmission across the DM channel may undergo one of three transitions. 1) With a probability of $P_i$, a bit may randomly be inserted into the received sequence. For $m$ consecutive insertions the probability is given as $P_i^m$. Following an insertion or multiple insertions, a transmission or deletion must follow to proceed to the next time step. Since, in theory, an unlimited number of consecutive insertions may occur, a maximum number of insertions, $I_m$, is imposed on the system for simplification. 2) A bit may be deleted from the sequence and will not appear in the received string with a probability $P_d$. 3) A bit may be transmitted with a probability $P_t$,  where {$P_t = 1 - P_i - P_d$} \cite{davey2000error,mackay,leigh}. Note that when $I_m$ insertions occur, a bit is transmitted with probability $\hat{P}_t = 1 - P_d$ \cite{gaurav2008insertion}. If a bit is transmitted, it may undergo a substitution with a probability $P_s$. The DM channel model is better illustrated in Figure \ref{fig:DMChannel} \cite{wang2012coding} where $\tau_n$  and $\tau_{n+1}$ indicates the time at $n$ and $n+1$ respectively.

The received sequence \textbf{r} is produced at the channel's output. Since the channel produces a mixture of insertions and deletions, the length of \textbf{r} is not necessarily equal to the length of \textbf{t}. The received sequence is then processed through an inner decoder where the goal is to produce a symbol-by-symbol likelihood function: $P_n(a) = P(\mathbf{r}|\mathbf{d_n}=a,P_i,P_d,P_s,P_t,f,\mathbf{w})$ \cite{gaurav2008insertion}. This involves making use of the Forward-Backward (FB) algorithm to infer the hidden states, which in this case is the synchronisation drift or the number of insertions minus the number of deletions the channel has made from the beginning of the channel use till time $\tau_n$ where the $n$th bit is queued to enter the channel. This FB algorithm is run firstly at a bit level, and a final forward pass is rerun at a symbol level. This produces a symbol-by-symbol likelihood which can be used as the input into the LDPC decoder. The LDPC decoder is a probabilistic iterative decoder that seeks to determine the marginal posterior probabilities for the codeword symbols.

The DM code construction has been successfully used in many applications ranging from the barcoding of DNA in \cite{kracht2014using,kracht2015insertion} to speech watermarking in \cite{gaurav2008insertion}. There has also been much work on improving the code construction and decoding, such as \cite{briffa2010improved,jiao2011interleaved,jiao2012soft}. These references still, however, establish the core channel model on the memoryless interpretation. 
 
\subsection{FSMC Synchronisation Channel}
\begin{figure}[th]
\centering
\includegraphics[width=\columnwidth]{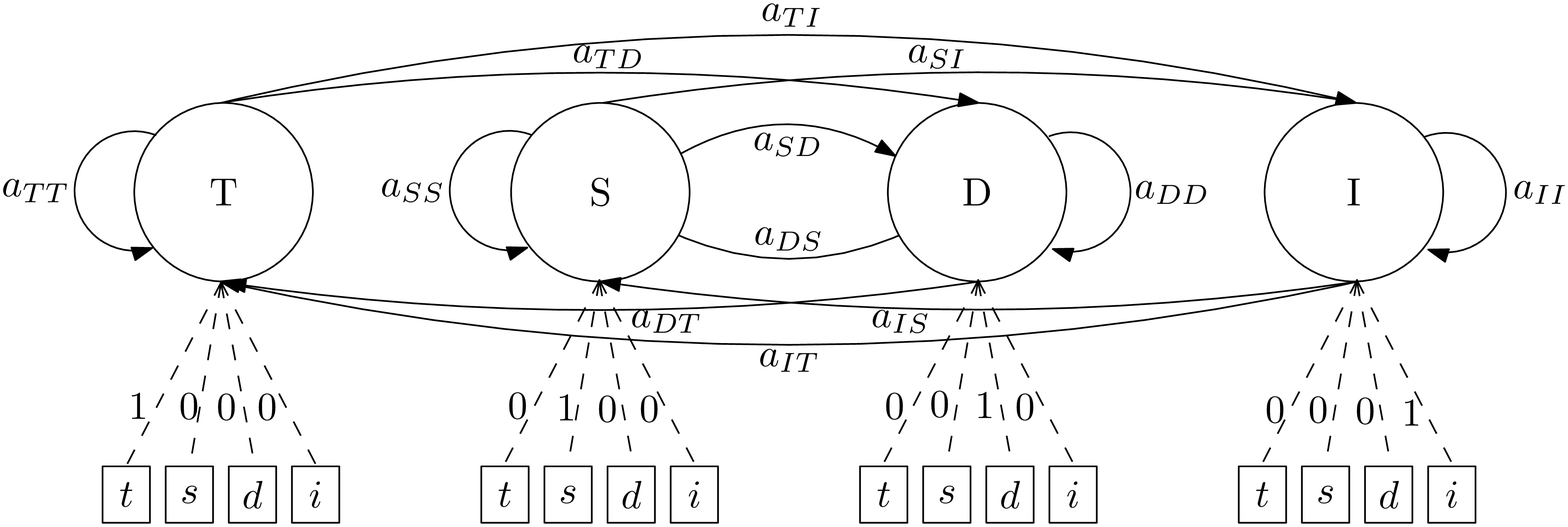}
\caption[Four-state FSMC for IDS channel]{Four-state FSMC for IDS channel\cite{achari2021symbol}}
\label{fig:FSMC}
\end{figure}

Achari et al. describe a method of creating a channel model for an IDS channel that contains memory and thus has correlations between errors \cite{achari2021symbol}. This contrasts with the DM synchronisation channel, which allows insertions, deletions and substitution errors to occur in an IID (essentially memoryless) manner. The model presented in \cite{achari2021symbol} is based on a FSMC where the states of the model are Transmission (T), Substitution (S), Deletion (D) and Insertion (I). Here, one may move between any two states, including self-transitions, with some probability defined in a corresponding transition matrix. The model is better illustrated in Figure \ref{fig:FSMC}. In \cite{achari2021symbol} the receiver is assumed to have full knowledge of the transmitted data. This allows the use of the Levenshtein Distance (LD or edit distance) algorithm to find the most likely series of events (transmission, substitution, deletion or insertion) that occurred to produce the final received string. Since this sequence is essentially the hidden states of the channel, the system is reduced to a Markov chain and the probabilities of the transition matrix are determined by using the Baum-Welch algorithm. The emission matrix, in this case, is reduced to an identity matrix with the dimensions equivalent to the number of states. The power of this modelling methodology is then shown by utilising data from a real-world visible light communication system and producing the corresponding channel model. As stated previously, this approach is extremely useful in creating the channel model and parameters. However, providing an encoding and decoding scheme to prevent such errors and resynchronise the data proves somewhat complicated.  

\section{Proposed Memory Synchronisation System}
\label{sec:Proposedsystem}

\subsection{Proposed Memory Synchronisation Channel}
\begin{figure}[th]
\centering
\includegraphics[width=0.9\columnwidth]{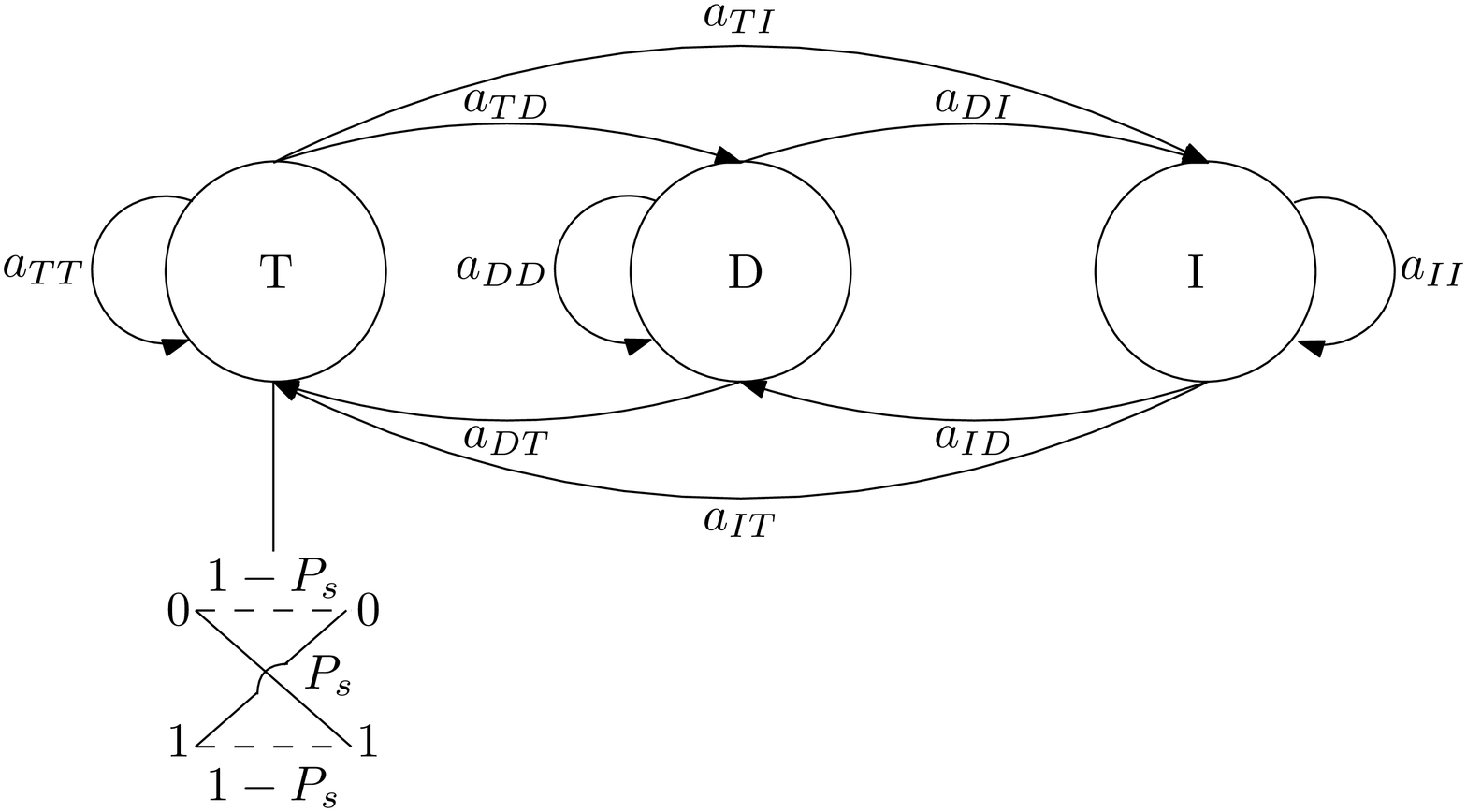}
\caption[Proposed three-state Markov model for synchronisation channel]{Proposed three-state Markov model for synchronisation channel}
\label{fig:proposedCM}
\end{figure}

This paper draws inspiration significantly from DM watermark codes \cite{davey2000error,mackay} and adapts the FSMC  in \cite{achari2021symbol} accordingly. In fact, this new proposed model can be viewed as an amalgamation of the DM model and the FSMC. Firstly, since the main timing and resynchronisation capabilities rely heavily on the watermark sequence, we need to ensure that when the system transitions from time $\tau_n$ to $\tau_{n+1}$ that a bit queued on the transmitter end actually enters the channel. This differs from the FSMC as time progresses to the next time step even if no transmitted bit is sent across the channel, e.g. insertion to another insertion will cause a time increment for the FSMC. While the new proposed model allows for multiple insertions, before moving to the next time step, either a deletion of the actual transmitted bit or transmission must occur, which is similar to the DM channel model. Another alteration from the FSMC is that a substitution may only occur if a transition occurs. Again, this is similar to the DM channel, and the probability of substitution is treated as IID. The new model is adapted only to include the transmission, deletion and insertion states and a corresponding memoryless probability of substitution. We can obtain this new three-state FSMC from the original four-state FSMC. Firstly, the stationary distributions of the four-state transition matrix is determined which provides the IID probabilities for $P_t$, $P_s$, $P_d$ and $P_i$. These probabilities are used to create a corresponding DM code construction and watermark decoding for comparative purposes and benchmarking. The value of $P_s$ is used in both the DM construction and as the corresponding memoryless substitution probability for the new proposed model. After obtaining the stationary distribution values, the columns and rows corresponding to substitutions in the four-state transition matrix are removed and normalised across the rows to obtain the three-state transition matrix. 
To simplify the situation further, a limit of $I_m$ maximum consecutive insertions is imposed on the channel - as with the DM model. If the maximum number of insertions is reached, the system must undergo a transition or a deletion and move to the next time step. This requires the value of insertion to insertion in the transition matrix to become zero (and normalise across the row) for this step. In all further simulations within this paper, the number of maximum insertions is capped to 1, which greatly simplifies the scenario. However, in theory, this may be restricted to any number - restricting this to zero produces a deletion channel. The proposed channel model is depicted in Figure \ref{fig:proposedCM} and the corresponding transition matrix, $A$, is given in Equation \ref{eqn:proposedTmat}. Accompanying the transition matrix is the initial distribution vector, $\Pi = \{\pi_{-x_{max}},\pi_{-x_{max+1}},...,\pi_{x_{max}}\}$, which represents the probabilities of starting in a particular state where $x_{max}$ is the maximum offset considered.
%Since the FSMC inherently contains an aspect of memory, we now look over the previous two time steps to produce the current state of the system - this is further clarified in the decoding of the proposed system.

\begin{equation}
\label{eqn:proposedTmat}
A = 
\begin{bmatrix} 
a_{TT} & a_{TD} & a_{TI}\\
a_{DT} & a_{DD} & a_{DI}\\
a_{IT} & a_{ID} & a_{II}\\
\end{bmatrix}
\quad
\end{equation}

\subsection{Proposed Code Construction}
\begin{figure}[th]
\centering
\includegraphics[width=0.9\columnwidth]{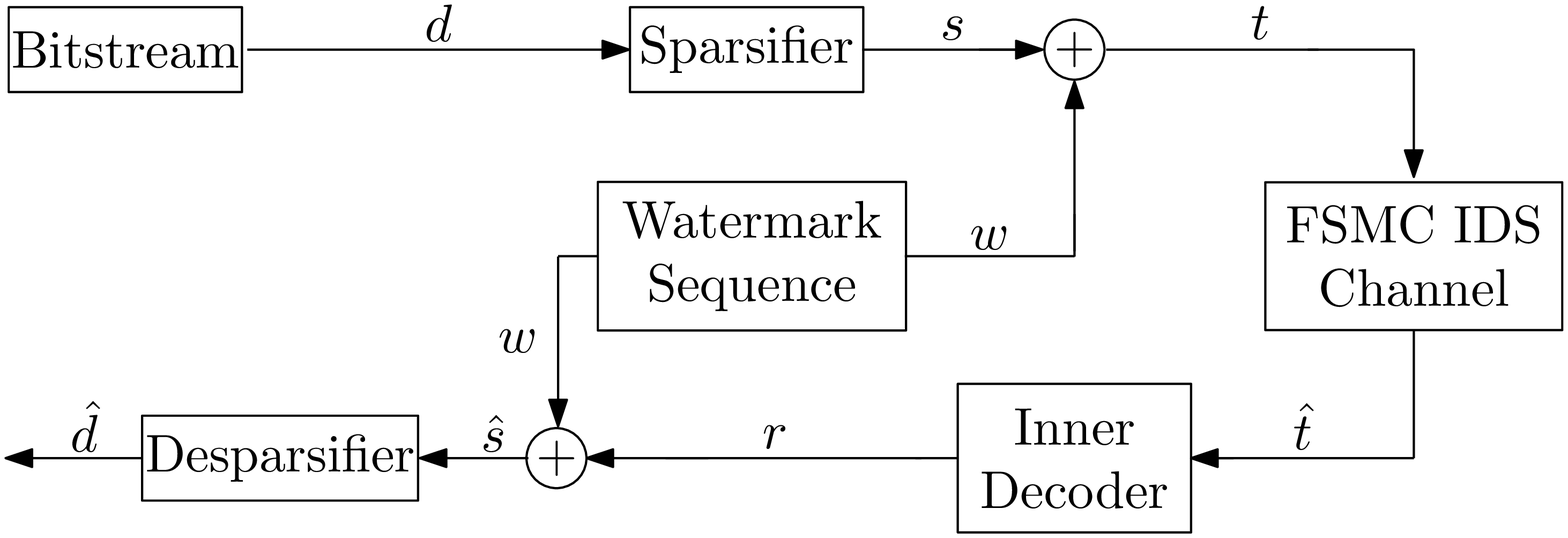}
\caption[Block diagram for system code construction]{Block diagram for system code construction}
\label{fig:proposedConstruct}
\end{figure}

The code construction for this system model once again draws substantially from DM watermark codes construction. In the proposed system, the main idea of the watermark scheme is used to regain synchronisation - the primary differences occur at the FSCM channel and Inner Decoder blocks in Figure \ref{fig:proposedConstruct} which illustrates the block diagram of the systems code construction. For this paper, the scope remains entirely on the inner decoder at a bit level to regain synchronisation. This is an added benefit of the proposed methodology and construction as the system is entirely independent of the outer encoder and decoder. As such, any substitution error correction code may be used as the outer code in conjunction with the presented inner code construction, and the system will not be restricted to using LDPC as in the original DM watermark scheme. Here the bitstream block represents the data in binary or the message, $d$. This data sequence is then processed through a sparsifier. As in the DM code construction, the sparsifier provides the decoder with its capabilities as only information bits and errors will likely produce a $1$ in the resulting received sequence. For simulations in this paper, a 4-to-5 sparsifier is used where the output code bits correspond to the sparsest permutations possible. The density of this output codebook and consequently the mean density of the output sequence of the sparsifier is again denoted by $f$. This sparse string, $s$, of length $\Gamma$ is then modulo two added to a watermark vector, $w$, (also of length $\Gamma$) to produce an encoded bitstream, $t$, ready to be sent over the channel. The channel's output is the received sequence, $\hat{t}$. This received sequence then passes through the (inner) decoder, which seeks to resynchronise the bitstream and remove any insertion or deletion errors from $\hat{t}$. The process and workings of the inner decoder are further detailed in Section \ref{sec:innerdec}. The resynchronised string, denoted by $r$, then has the watermark sequence removed from it to produce $\hat{s}$ which is the sparse decoded sequence. The sparse decoded sequence is then fed into a desparsifier which reverses the operations of the sparsifer to produce $\hat{d}$, which is the received data bitstream. It is worth noting that the resynchronisation process ultimately introduces additional substitution errors into the sequence that were not caused by the channel. Such a system will undoubtedly benefit from an (outer) encoder and decoder.

\subsection{Inner Decoder}
\label{sec:innerdec}
The purpose of the inner decoder is to regain synchronisation by accounting for the errors caused by insertions and deletions. A probabilistic approach is used to determine the most likely hidden states the channel traversed to produce the given received string. This decoding is based on Hidden Markov Model (HMM) decoding, and in particular, we use the FB algorithm on a bit level. For more information regarding standard HMMs and their respective decoding, readers are encouraged to review \cite{rabiner,hmmch8}. What we present is not an HMM in the traditional sense, and as indicated in \cite{davey2000error,mackay}, there are some subtle variations. Firstly the hidden states of the HMM are not the same as the states of the transition matrix. Here the hidden state is defined as the synchronisation offset or synchronisation drift (number of insertions minus number of deletions) at a given time instance. As in the DM code construction, we also restrict the maximum offset to $x_{max}$ to simplify the calculations as our number of states ranges from $-x_{max}$ to $x_{max}$. In our case, $x_{max}$ is equal to five times the absolute final offset. For example, if 600 bits were transmitted and only 595 bits were received, the final offset, $\Psi$, at the final time, $\Gamma$, would be $-5$ and the range of possible hidden states would be $-25$ to $25$. The exception is when $\Psi$ is zero (equal number of insertions and deletions occurring during data transmission), then $x_{max}$ is set to 5. Increasing this range allows for a truer representation of the channel and could potentially increase accuracy, but this comes with a trade-off of increasing the number of calculations. The second alteration is that the transition matrix may not be used directly in the FB calculations, and the transitions are instead based on channel events. To better illustrate the transition between each time step, we use the number of bits output by the channel at a given time interval. Since two time intervals (three time steps) are used, we indicate the offset at times $\tau_{n-2}$, $\tau_{n-1}$ and $\tau_{n}$  as $\psi_{n-2} = i$, $\psi_{n-1} = j$ and $\psi_{n} = k$ respectively. At each time interval ($\tau_{n-2}$ to $\tau_{n-1}$ or $\tau_{n-1}$ to $\tau_{n}$) the channel may output from zero to $I_m+1$ bits depending on the events of the channel. This bit output and corresponding channel events are better illustrated in Table \ref{tbl:bitoutput}.

\begin{table}
\centering
\caption{Bits Output by Channel and Corresponding Channel Event(s)}
\label{tbl:bitoutput}
\resizebox{\linewidth}{!}{%
\begin{tabular}{cc} 
\toprule
\multicolumn{1}{c}{Bits Output by Channel} & \multicolumn{1}{c}{Channel Event}                                                                                 \\ 
\midrule
0                                          & Deletion                                                                                                          \\ 
\hline
1                                          & \begin{tabular}[c]{@{}c@{}}Transmission\\OR \\(Insertion and Deletion)\end{tabular}                               \\ 
\hline
2                                          & \begin{tabular}[c]{@{}c@{}}(Insertion and Transmission) \\OR \\(Two Insertions and Deletion)\end{tabular}         \\ 
\hline
3                                          & \begin{tabular}[c]{@{}c@{}}(Two Insertions and Transmission) \\OR \\(Three Insertions and Deletion)\end{tabular}  \\ 
\hline
\vdots                                      & \vdots                                                                                                           \\ 

$I_m + 1$                                    & ($I_m$ Insertions and Transmission)                                                                                 \\
\bottomrule
\end{tabular}
}
\end{table}

From the number of bits output by the channel, and consequently the events that produced such an output, various transition probabilities that are based on the three-state transition matrix can be derived. Table \ref{tbl:correspondingprob} illustrates the probabilities used for the various bit emissions over two-time intervals. It is worth noting that values in this table correspond to $I_m=1$ and the table will differ depending on this parameter. The number of possibilities also increases as $I_m$ increases as more potential events are now possible. Additionally, when the final event contains an insertion we scale the value over all possible bits i.e $2^m$ where $m$ once again corresponds to $m$ consecutive insertions. Note that the above notation is for the forward pass of the FB algorithm and the corresponding notation for the backward pass describes $\psi_{n} = k$, $\psi_{n+1} = j$ and $\psi_{n+2} = i$  as the synchronisation offset at $\tau_n$, $\tau_{n+1}$ and $\tau_{n+2}$ respectively. Additionally, the time interval from $\tau_{n-2}$ to $\tau_{n-1}$ in Table \ref{tbl:correspondingprob} corresponds to time interval from $\tau_{n+1}$ to $\tau_{n+2}$ for the backward pass and similarly time interval from $\tau_{n-1}$ to $\tau_{n}$ will correspond to time interval from $\tau_{n}$ to $\tau_{n+1}$. Likewise $P_{ijk}$ in Table \ref{tbl:correspondingprob} corresponds to $P_{kji}$ for the backward pass.

\begin{table}
\centering
\caption{Bits Output by Channel and Corresponding Probabilities}
\label{tbl:correspondingprob}
\resizebox{\linewidth}{!}{%
\begin{tabular}{ccc} 
\toprule
\begin{tabular}[c]{@{}c@{}}Bits Output\\($\tau_{n-2}$ to $\tau_{n-1}$)\end{tabular} & \begin{tabular}[c]{@{}c@{}}Bits Output\\($\tau_{n-1}$ to $\tau_{n}$)\end{tabular} & \begin{tabular}[c]{@{}c@{}}Probability\\($P_{ijk}$)\end{tabular}  \\
\midrule
0     & 0                                                             & $a_{DD}$     \\
0     & 1                                                             & $a_{DT} + a_{DI}\frac{a_{ID}}{2}$ \\
0     & 2                                                             & $a_{DI}\frac{a_{IT}}{2}$ \\
1     & 0                                                             & $a_{TD} + a_{ID}a_{DD}$ \\
1     & 1                                                             & $a_{TT} + a_{TI}\frac{a_{ID}}{2}+ a_{ID}a_{DT} + a_{ID}a_{DI}\frac{a_{ID}}{2} $ \\
1     & 2                                                             & $a_{TI}\frac{a_{IT}}{2} + a_{ID}a_{DI}\frac{a_{IT}}{2}$ \\
2     & 0                                                             & $a_{IT}a_{TD}$             \\
2     & 1                                                             & $a_{IT}a_{TT} + a_{IT}a_{TI}\frac{a_{ID}}{2}$             \\
2     & 2                                                             & $a_{IT}a_{TI}\frac{a_{IT}}{2}$             \\
\bottomrule
\end{tabular}
}
\end{table}

The final alteration is that there is no emission matrix for an observation given a current state. Rather, we use the watermark sequence and compare corresponding received bits to determine if a transmission or substitution has occurred. Again this shows the importance of the watermark sequence and the density of the sparse sequence. An altered FB algorithm is used to determine the likely states that the channel transitioned through. The forward probabilities at time $n$ and state $k$ is defined as $F_n(k) = P(\hat{t}_1,\hat{t}_2,...,\hat{t}_{n-1+k}, \psi_n = k | A,P_s,w,\Pi)$ \cite{gaurav2008insertion}. Similarly, the backward probabilities at time $n$ for state $k$ is defined as $B_n(k) = P(\hat{t}_{n+k},\hat{t}_{n+k+1},...,\hat{t}_{\Gamma}, \psi_n = k | A,P_s,w)$ \cite{gaurav2008insertion}. As mentioned previously, an efficient method to calculate these probabilities is with the recursive FB algorithm. The equations for the forward pass are shown in Equations \eqref{eqn:Forward1} to \eqref{eqn:Forwardr} and the backward pass formulae are shown in Equations \eqref{eqn:BackwardT} to \eqref{eqn:Backwardr}. Additionally, Equation \eqref{eqn:Forward1} is used for the initialisation step and Equation \eqref{eqn:Forward2} is used for the recursion in the first-order memoryless DM decoding. Similarly, Equation \eqref{eqn:BackwardT} is used for the initialisation step and Equation \eqref{eqn:BackwardT1} is used for the recursion in the backward pass for the first-order memoryless DM decoder. To better illustrate the processes described, the algorithm for the FB passes for the proposed decoding scheme is outlined in Algorithm \ref{alg:FBFSMC} in \ref{app:algorithms}. The probability of being in a given state at a given time, or the posterior state probabilities, is equal to the product of the forward and backward values at the same corresponding state and time. From these posterior state probabilities, the most likely channel path sequence is determined by finding the state with the maximum probability for each time index constrained within a range of $s_p-1$ to $s_p+I_m$ where $s_p$ is the most likely state at the previous time index.
\\ \\

for $n=1$

\begin{equation}
\label{eqn:Forward1}
F_1(k) = \pi_k
\end{equation}

for $n=2$ 

\begin{equation}
\label{eqn:Forward2}
F_2(k) =  \sum_{j=k-I}^{k+1} F_1(j) \left(\alpha_{jk} + \beta_{jk}\zeta_{k}^{1}\right)
\end{equation}

for $3 \leq n \leq \Gamma$

\begin{equation}
\label{eqn:Forwardr}
F_n(k) =  \sum_{i=j-I}^{j+1} \sum_{j=k-I}^{k+1} F_{n-1}(j) P_{ijk} \xi_{k}^{n-1}
\end{equation}

%%%%%%%%%%%%%%%%%%%%%%%%%%%%%%%%%%%%%%%%%%%%%%%%%%%%%%%%%%%%%%%%%
for $n=\Gamma$

\begin{equation}
\label{eqn:BackwardT}
B_\Gamma(k) = \left\{
\begin{array}{ll}
      1 & k=\rho \\
      0 & otherwise\\
\end{array} 
\right.
\end{equation}

for $n=\Gamma-1$

\begin{equation}
\label{eqn:BackwardT1}
B_{\Gamma-1}(k) =  \sum_{j=k-1}^{k+I} B_\Gamma(j) \left(\alpha_{kj} + \beta_{kj}\zeta_{k}^{\Gamma-1}\right)
\end{equation}

for $\Gamma-2 \geq n \geq 1$

\begin{equation}
\label{eqn:Backwardr}
B_n(k) =  \sum_{i=j-1}^{j+I} \sum_{j=k-1}^{k+I} B_{n+1}(j) P_{kji} \xi_{k}^{n}
\end{equation}

Here $\alpha_{jk}$, $\beta_{jk}$ and $\zeta_{k}^{n}$ are further elaborated in Equations \eqref{eqn:alpha},\eqref{eqn:beta} and \eqref{eqn:zeta} respectively \cite{gaurav2008insertion}. $\xi_{k}^{n}$ in Equations \eqref{eqn:Forwardr} and \eqref{eqn:Backwardr} are equivalent to $\zeta_{k}^{n}$ in Equation \eqref{eqn:zeta}, however, it only affects the probability in question if the second time interval ends in a Transmission. For example, the calculation for $\psi_{n-2} = 1$, $\psi_{n-1} = 0$ and $\psi_{n} = 0$, the channel emits 0 bits for the time interval $\tau_{n-2}$ to $\tau_{n-1}$ and 1 bit for $\tau_{n-1}$ to $\tau_{n}$. This means that the overall $P_{ijk} \xi_{k}^{n-1}$ for this case would be $a_{DT}\xi_{0}^{n} + a_{DI}\frac{a_{ID}}{2}$. In other words the value of $\xi_{k}^{n-1}$, which checks if the received bit matches the corresponding watermark bit, is only used if there was in fact a transmission event as the final respective occurrence. 

\begin{equation}
\label{eqn:alpha}
\alpha_{jk} = \left\{
\begin{array}{ll}
      \frac{P_i^{k-j+1}P_d}{2^{k-j+1}}     & -1 \leq k-j < I\\
      0                                    & k-j <-1, k-j \geq I\\
\end{array} 
\right.
\end{equation}

\begin{equation}
\label{eqn:beta}
\beta_{jk} = \left\{
\begin{array}{ll}
      \frac{P_i^{k-j}P_t}{2^{k-j}}     & 0 \leq k-j < I\\
      \frac{P_i^{I}\hat{P}_t}{2^{k-j}}     & k-j  = I\\
      0                                    & k-j \leq -1, k-j > I\\
\end{array} 
\right.
\end{equation}

\begin{equation}
\label{eqn:zeta}
\zeta_{k}^n = \left\{
\begin{array}{ll}
      1-P_f     & \hat{t}_{n+k} = w_n\\
      P_f     & \hat{t}_{n+k} = w_n \oplus 1\\
\end{array} 
\right.
\end{equation}

\section{Simulations and Results}
\label{sec:Simulations}
As previously mentioned, all simulation results from the proposed model and code construction are benchmarked against the first and second-order memoryless DM decoding. Algorithm \ref{alg:FBDM1} in \ref{app:algorithms} shows the pseudo-code for the first-order DM FB algorithm. We expand the first-order decoder to a second-order, providing a more objective comparison to the proposed scheme as two-time intervals are now used to evaluate the forward and backward values. Equation \eqref{eqn:Forward2Or} provides the memoryless second-order recursive forward pass while Equation \eqref{eqn:Backward2Or} describes the memoryless second-order recursive backwards pass. The initialisation steps follow the same process as the proposed FSMC decoder. The second-order DM FB algorithm pseudo-code is given in Algorithm \ref{alg:FBDM2} in \ref{app:algorithms}.

%%%%%%%%%%%%%%%%%%%%%%%%%%%%%%%%%%%%%%%%%%%%%%%%%%%%%%%%%%%%%%%%%%%%%%%%%%%%%%%

\begin{equation}
\label{eqn:Forward2Or}
F_n(k) =  \sum_{i=j-I}^{j+1} \sum_{j=k-I}^{k+1} F_{n-1}(j) \left(\alpha_{ij} + \beta_{ij}\right) \left(\alpha_{jk} + \beta_{jk}\zeta_{k}^{n-1} \right)
\end{equation}

\begin{equation}
\label{eqn:Backward2Or}
B_n(k) =  \sum_{i=j-1}^{j+I} \sum_{j=k-1}^{k+I} B_{n+1}(j) \left(\alpha_{ji} + \beta_{ji}\right) \left(\alpha_{kj} + \beta_{kj}\zeta_{k}^{n} \right)
\end{equation}

\subsection{Analysis Metrics}
Three evaluation metrics are used in the analysis of the simulation results. The first is the commonly used BER values for a given system entropy. While this provides a good insight into the system's performance, there are some minor drawbacks. The BER is calculated at the final output of the system; in other words, $\hat{d}$ is compared against $d$ to determine the equivalent BER. This requires a complete resynchronisation of the received sequence. Consequently, if a deletion is detected during the inner decoding process, a bit is inserted into the received sequence at the corresponding time index to rectify this influence. In real applications, the rectified bit is random and may sometimes produce the correct bit that was deleted. Other times this may be the incorrect bit, which will alter the BER values and thus produce varying results for the different algorithms. For the analysis in this paper, the bit inserted to undo a possible deletion is always a '0' to ensure a fair comparison of the BER analysis for the different decoders. The bias caused is still, however, noted in this resynchronisation step which ultimately biases the algorithms BER performance.
The second evaluation metric used is the number of positions that the derived hidden states disagree with the actual states that the channel traversed - we call this the NIIS or Number of Incorrectly Identified States. This indicates how well the respective FB algorithms could accurately deduce the correct hidden states. However, it suffers from the fact that the correct transitions between states may be correct, but the actual state (offset) is incorrect. Here a lower number indicates a more accurate representation of what truly transpired in the channel. For the results presented in this paper, the NIIS is normalised over the length of the transmitted sequence in order to provide a more general metric.
Finally, an evaluation metric known as the Sum of Absolute Offset ($SAO$), which builds onto the idea of NIIS, is introduced. This metric is derived again at the inner decoder and looks at how the likely hidden state path produced from the algorithms differs from the actual path the channel traversed. Instead of looking at the number of individual states where the algorithm path matches the actual path (binary classification), we look at the difference between the actual path and algorithm-derived path at corresponding time indexes. This gives a more general overview and a value that better shows the performance of the associated algorithm where, again, the lower the $SAO$, the more aligned to the actual channel path. The $SAO$ is calculated by Equation \eqref{eqn:SAO} where $P_{Actual}(\tau)$ and $P_{Alg}(\tau)$ are the actual state path produced by the channel and the derived state path inferred from the decoding algorithms respectively. In this paper, the results for the $SAO$ plots are kept as absolute values as they closely correlate to what is observed in the NIIS, and this allows us to have a different perspective when viewing the results. It is noted that the $SAO$ can, however, still be normalised across all possible offsets and the length of the transmitted sequence if needed.

\begin{equation}
\label{eqn:SAO}
SAO =  \sum_{n=1}^{\Gamma} |P_{Actual}(n) - P_{Alg}(n)|
\end{equation}

These evaluation metrics are plotted against a range of entropy values in the results. Entropy is used as a measure of how uncertain the communication channel is. In other words, we use entropy to quantify the harshness of the channel as it is a good indicator of channel capacity \cite{kanal1978models}. The average entropy is calculated by Equation \eqref{eqn:FMent} where the entropy of state $i$ is $H_i$ and is calculated by Equation \eqref{eqn:FMHi} \cite{shanmugam1979digital, achari2021symbol}. The value of $\rho_i$ in Equation \ref{eqn:FMent} is the stationary distribution of been in that specific state. As different combinations of $P_t$, $P_s$, $P_d$, $P_i$ and $A$ may produce similar entropies, the justification of using entropy in this regard is to limit the number of these varying parameters in the analysis, which seeks to simplify a rather complex phenomena. As can be seen in the overall results, having varying parameters within certain ranges reliably reproduces relatively constant entropies. In this paper, three error ranges are used for the generation of the transition matrix entries, and this, in turn, corresponds to three different entropy ranges. These ranges are shown in Table \ref{tbl:transprob} in \ref{app:transmatcreation} along with the outline and method for creating the transition matrices. We, however, reiterate that the objective of this paper is to give a comparison of the decoding algorithms. While this matter is briefly discussed, the manner in which the different parameters contribute to the entropy is beyond the scope of this paper. 

\begin{equation}
\label{eqn:FMent}
\overline{H} = \sum_{i=1}^{N} \rho_i H_i \quad \textrm{\textit{bits/symbol}}
\end{equation}

\begin{equation}
\label{eqn:FMHi}
H_i = -\sum_{j=1}^{N} a_{ij} \log_2(a_{ij})  \quad \textrm{\textit{bits/symbol}}
\end{equation}

\subsection{Overall Results}
The following results are obtained to show the general trends in performance of the three decoding algorithms across various entropy values. To ensure the generality and demonstrate the practicality of the decoders, for these tests, multiple iterations are run on varying transition matrices that produce an entropy within 0.001 from the desired entropy value. For example, if the desired entropy is 0.01, transition matrices that produce entropies between 0.009 to 0.011 are considered. Twenty different transition matrices are generated for each entropy value from 0.01 to 0.3, according to the method outlined in \ref{app:transmatcreation}. For each of these matrices, communication across the channel is simulated 100 times with varying message data, and from this, the decoding performances are computed. The results are then averaged across all iterations to produce the following graphs and outcomes.

\begin{figure}[tp]
\centering
\begin{subfigure}{\columnwidth}
\includegraphics[width=\columnwidth]{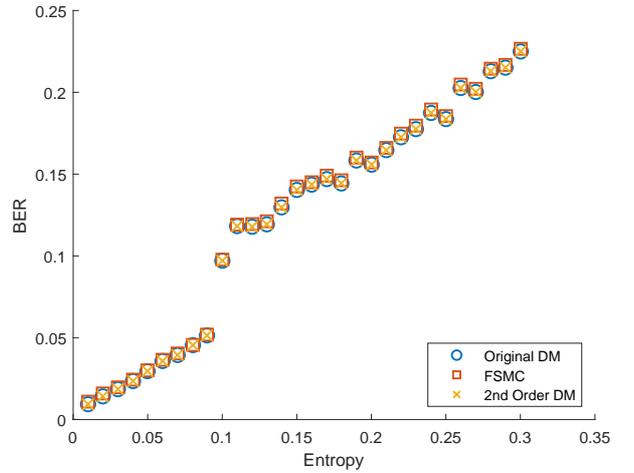}
\caption[BER performance for decoding algorithms]{BER performance for decoding algorithms}
\label{fig:overallBER}
\end{subfigure}
\hfill
\begin{subfigure}{\columnwidth}
\centering
\includegraphics[width=\columnwidth]{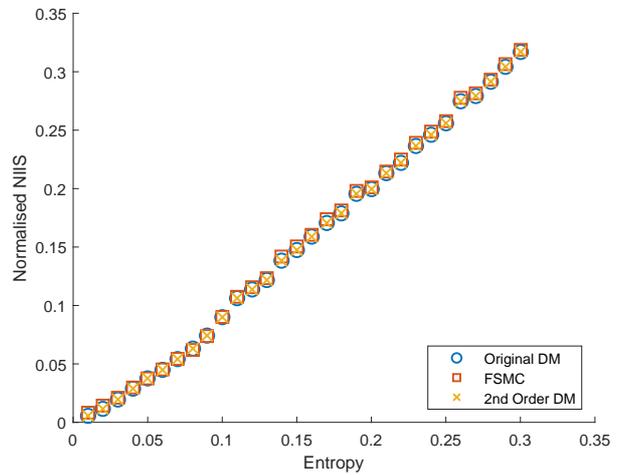}
\caption[NIIS performance for decoding algorithms]{NIIS performance for decoding algorithms}
\label{fig:overallNIIS}
\end{subfigure}

\begin{subfigure}{\columnwidth}
\centering
\includegraphics[width=\columnwidth]{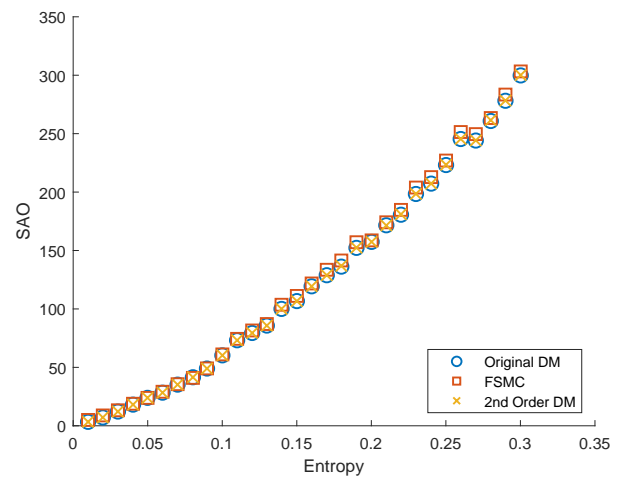}
\caption[SAO performance for decoding algorithms]{SAO performance for decoding algorithms}
\label{fig:overallSAO}
\end{subfigure}

\caption{BER, NIIS and SAO performance for the DM, FSMC and second-order DM decoding algorithms across various entropy values using varying transition matrices for each of the entropy values}
\label{fig:overallall}
\end{figure}

As shown in Figure \ref{fig:overallall}, all the evaluation metrics follow similar trends for the three decoding algorithms. As the entropy values increases so too does the BER, NIIS and $SAO$ which is shown in Figures \ref{fig:overallBER}, \ref{fig:overallNIIS} and \ref{fig:overallSAO} respectively. This is an expected outcome as an increase in entropy increases the uncertainty of what is occurring within the communication channel, and as a result, more errors are produced. These results agree with the notion of memory decreasing entropy and uncertainty and, consequently, increasing channel capacity  \cite{kanal1978models}. At lower entropy values, the memory FSMC outperforms its memoryless counterpart. As the entropy increases, the memoryless algorithms perform slightly better. For almost all the simulations, the first- and second-order memoryless DM algorithms perform identically, but the second-order decoder performance is slightly better at the much higher entropies. What is advantageous from these results is that we see entropy is indeed a good indicator of the channel's performance. Even though different transition matrices that produce a given entropy are used, the results obtained are generally independent of the individual entries of $A$ and the stationary distribution values. This is welcomed to demonstrate the real-world application of such a  system as the results are not constrained to specific entries of $A$ but can rather be overviewed by an encompassing metric like entropy. We also note jumps in the trajectory of the plots in Figure \ref{fig:overallBER} occurring at entropy values of $0.1$ and $0.2$. These discontinuities are attributed to the different error ranges outlined in Table \ref{tbl:transprob} in \ref{app:transmatcreation}. We propose that using a smaller resolution of changing error values will decrease this discontinuity, but this is left as a future exercise.

%%%%%%%%%%%%%%%%%%%% Total BER NIIS and SAO
\begin{figure}[H]
\centering
\begin{subfigure}{0.96\columnwidth}
    \includegraphics[width=\columnwidth]{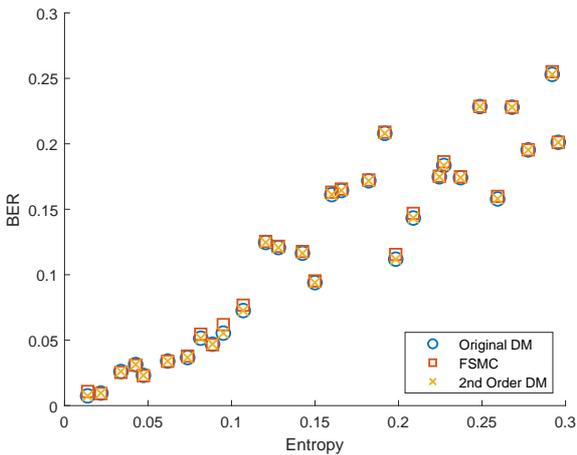}
\caption[BER performance for decoding algorithms]{BER performance for decoding algorithms}
\label{fig:BERtotal}
\end{subfigure}

\begin{subfigure}{0.96\columnwidth}
    \includegraphics[width=\columnwidth]{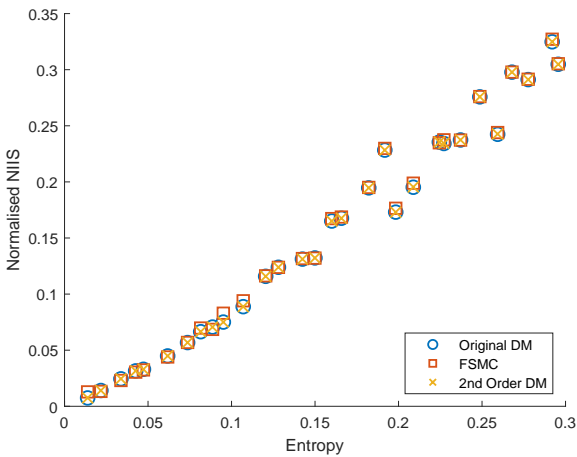}
\caption[NIIS performance for decoding algorithms]{NIIS performance for decoding algorithms}
\label{fig:NIIStotal}
\end{subfigure}
\caption{BER, NIIS and SAO performance for the DM, FSMC and second-order DM decoding algorithms across various entropy values}
\end{figure}

\begin{figure}[H]
 \ContinuedFloat

\begin{subfigure}{0.96\columnwidth}
    \includegraphics[width=\columnwidth]{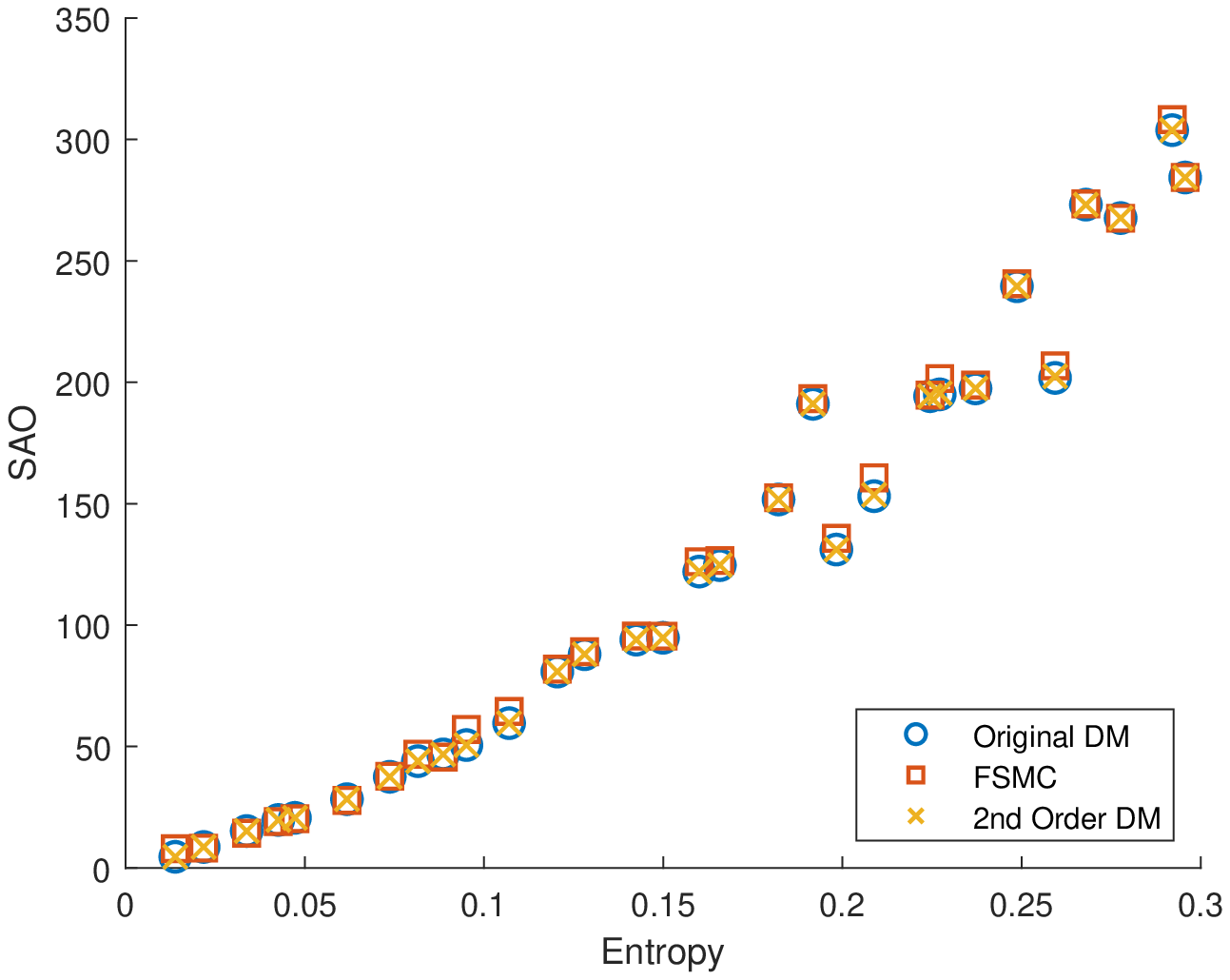}
\caption[SAO performance for decoding algorithms]{SAO performance for decoding algorithms}
\label{fig:SAOtotal}
\end{subfigure}

\caption{BER, NIIS and SAO performance for the DM, FSMC and second-order DM decoding algorithms across various entropy values (cont.)}
\label{fig:fig}
\end{figure}

\subsection{Constant Entropy Results}
\begin{figure}[th]
\centering
\includegraphics[width=\columnwidth]{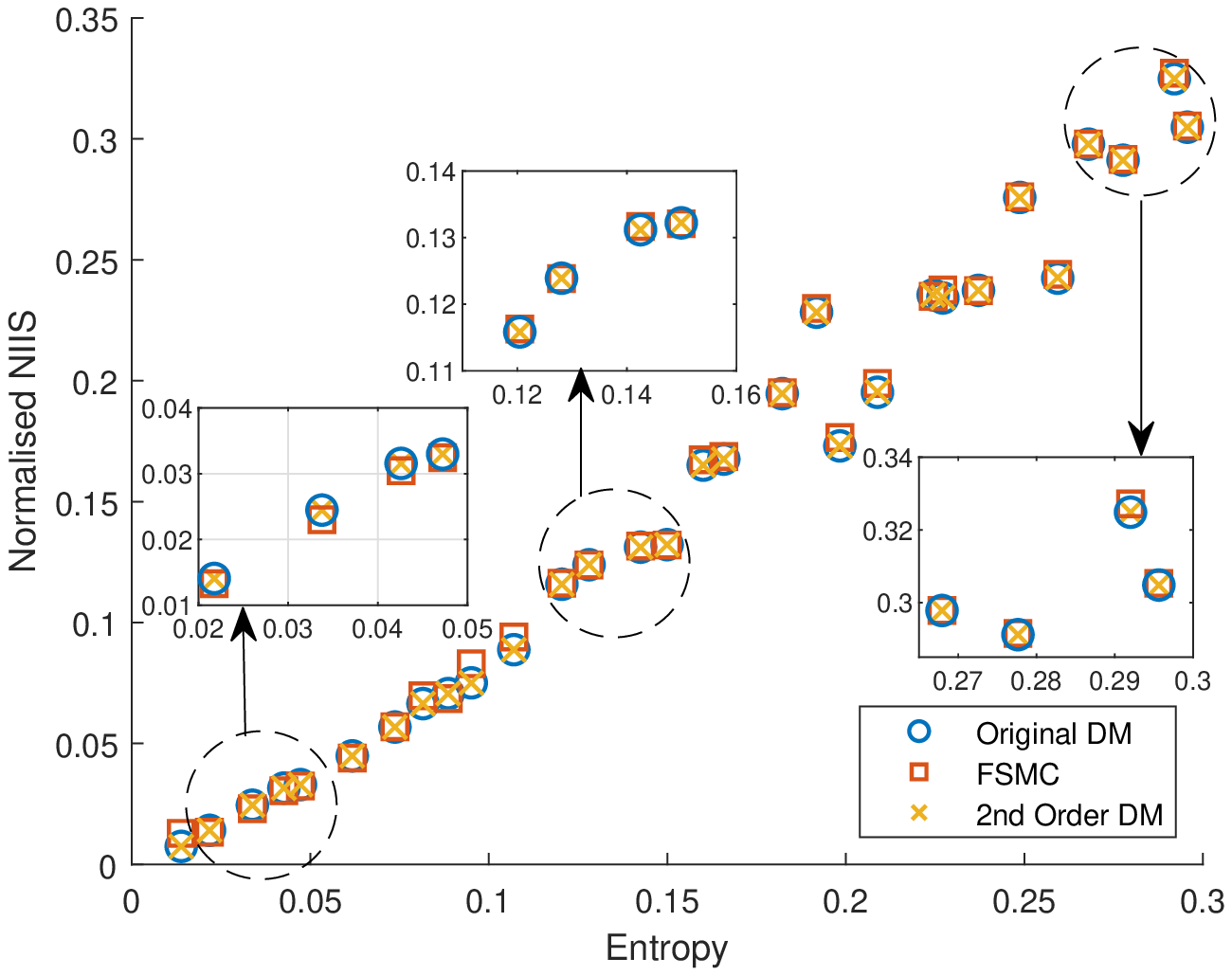}
\caption[NIIS performance for decoding algorithms with zoomed areas]{NIIS performance for decoding algorithms with zoomed areas}
\label{fig:NIISzoomed}
\end{figure}

The following results are obtained by running the three decoding algorithms on data passed through the channel for specific entropy values. Here a constant transition matrix and consequently constant entropy value within the desired range is used. This reduces the number of changing parameters, allowing a more thorough analysis of $P_s$, $P_d$ and $P_i$ values. For each entropy value, there are 5000 iterations of the channel use with random data bits for each run (all three algorithms are run using the same data and channel output for a true representation for the given iteration). For all entropies and iterations, the same watermark string is used to reduce the number of varying parameters. Figure~\ref{fig:BERtotal}, Figure~\ref{fig:NIIStotal} and Figure~\ref{fig:SAOtotal} show the results of these tests for the BER, NNA and $SAO$ respectively. The results obtained using set transition matrices for a given entropy range are in direct comparison to those obtained previously, where the transition matrix was varied for an entropy range. Additionally, it is noted that the results from the BER, NIIS and $SAO$ are highly correlated and in agreement with one another and as such, the focus can remain on the NIIS plots for further analysis. Figure~\ref{fig:NIISzoomed} shows specific zoomed-in areas of the NIIS plots to illustrate further the results obtained. From Figure~\ref{fig:NIISzoomed} it can be seen that using an entropy range from $0$ to $0.1$, the FSMC memory decoding appears to have slightly better performance than both the DM and second-order DM decoding. From an entropy range of $0.11$ to $0.2$, it is evident that all of the decoders have similar performance and achieve almost identical results. Lastly, for entropies ranging from $0.2$ to $0.3$, it is seen that both the DM decoders perform better than the FSMC in most cases. We also see the second-order DM decoding slightly outperforming the first-order DM algorithm at these higher entropy values. It is also worth noting that while the proposed model and scheme perform only slightly better, within lower entropy ranges, than the memoryless counterparts, there is almost no increase in the algorithm's complexity. Additionally, using the more sophisticated FSMC memory model allows for more realistic channels to be simulated and coded for. The emphasis is once again placed on providing an effective model and code construction to correct and protect against synchronisation errors in a channel that contains memory. It is easily seen that when the proposed scheme does not perform better than the existing approximations, it is at the very least comparable to them.

%%%%%%%%%%%%%%%%%%%% Broken up NIIS Plot

%\begin{figure}[tH]
%\begin{subfigure}{\columnwidth}
%\centering
%%\includegraphics[width=\columnwidth]{Figures/Results/NIIS1to10.eps}
%\caption[Block diagram for system code construction]{Block diagram for system code construction}
%\label{fig:proposedConstruct}
%\end{subfigure}

%\begin{subfigure}{\columnwidth}
%\centering
%\includegraphics[width=\columnwidth]{Figures/Results/NIIS10to20.eps}
%\caption[Block diagram for system code construction]{Block diagram for system code construction}
%\label{fig:proposedConstruct}
%\end{subfigure}

%\begin{subfigure}{\columnwidth}
%\centering
%\includegraphics[width=\columnwidth]{Figures/Results/NIIS21to30.eps}
%\caption[Block diagram for system code construction]{Block diagram for system code construction}
%\label{fig:proposedConstruct}
%\end{subfigure}

%\caption{Put your caption here}
%\label{fig:fig}
%\end{figure}

\subsection{Error Probability Level Results}
\begin{figure*}[h]
\centering
\setkeys{Gin}{width=\textwidth} %set image parameters
\begin{subfigure}{0.48\textwidth}
    \centering
    \includegraphics[width=\linewidth]{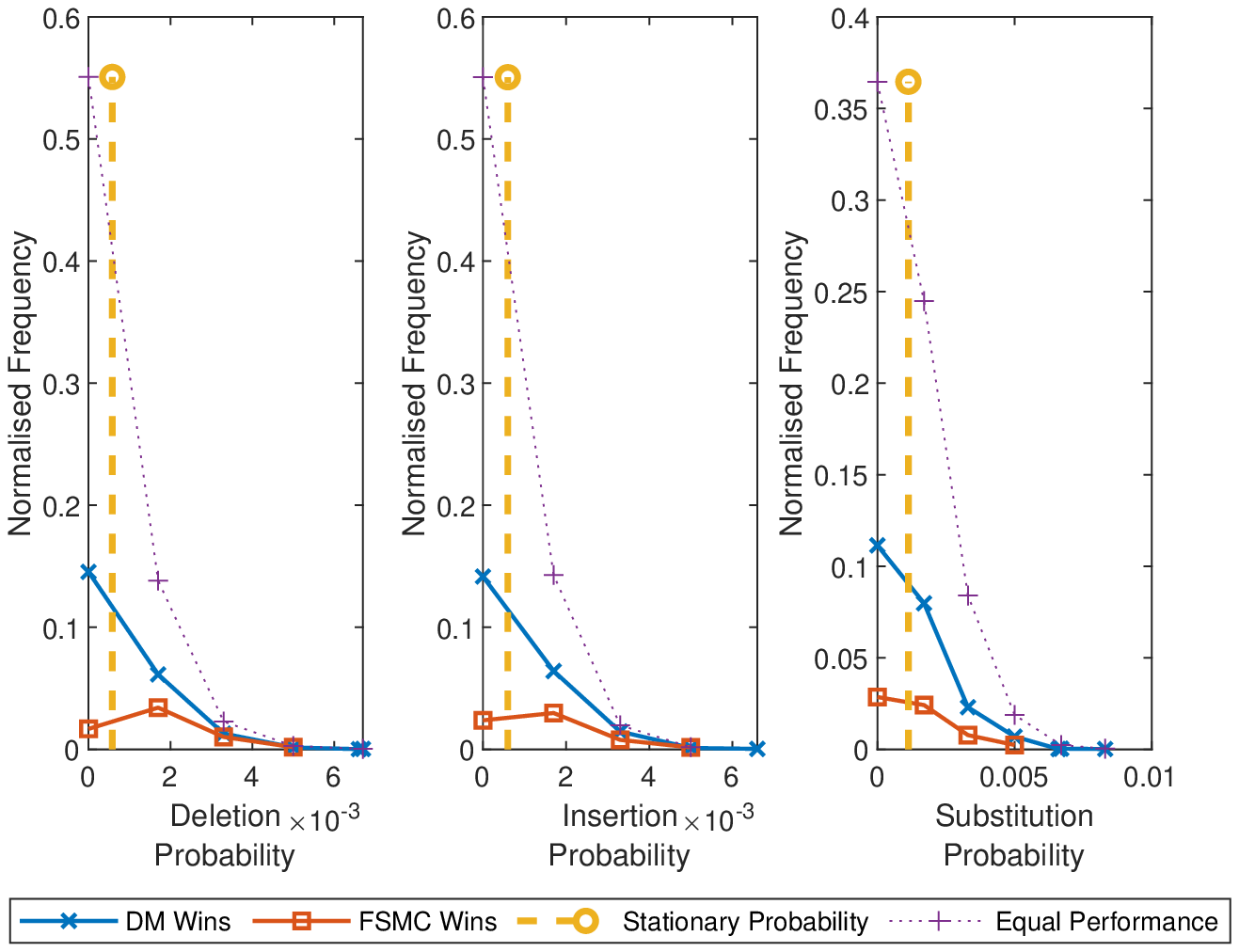}
        \caption[$H=0.014$]{$H=0.014$}    
        \label{fig:freq1}
\end{subfigure}
\hfil
\begin{subfigure}{0.48\textwidth}
    \centering
    \includegraphics[width=\linewidth]{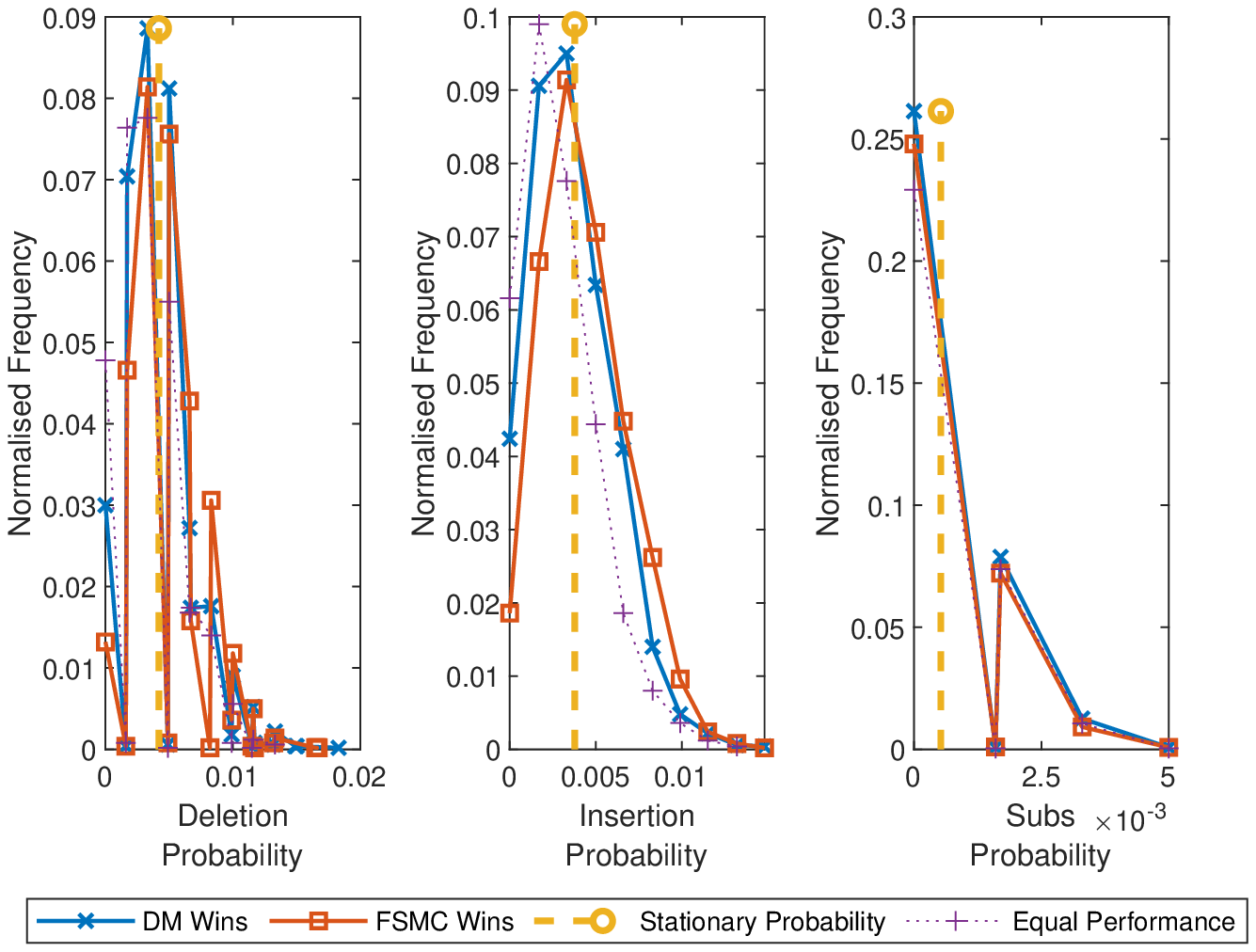}
    \caption[$H=0.074$]{$H=0.074$}    
    \label{fig:freq2}
    \end{subfigure}

    \begin{subfigure}{0.48\textwidth}   
    \includegraphics[width=\linewidth]{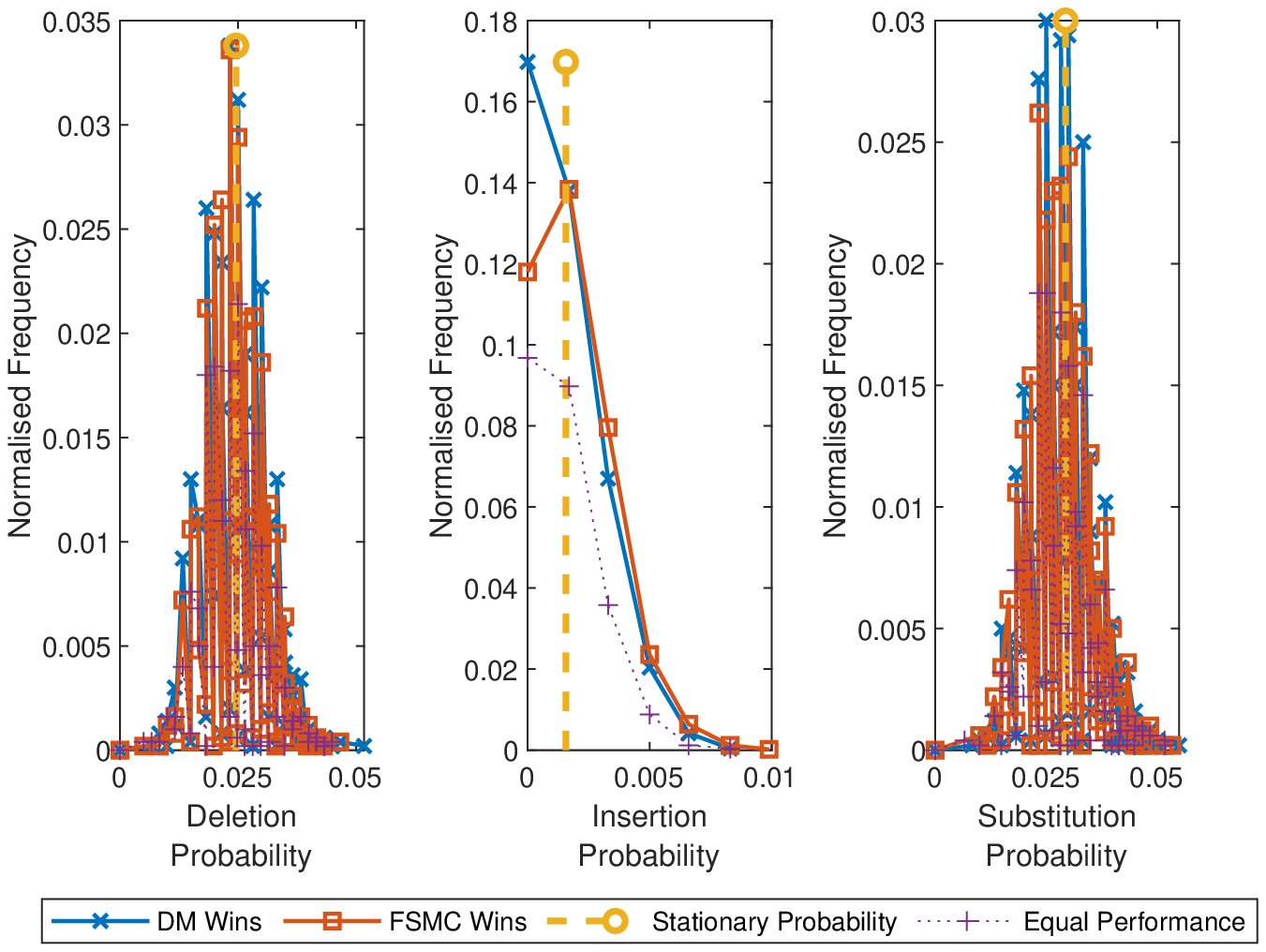}
    \caption[$H=0.182$]{$H=0.182$}    
    \label{fig:freq3}
    \end{subfigure}
\hfil
    \begin{subfigure}{0.48\textwidth}   
    \includegraphics[width=\linewidth]{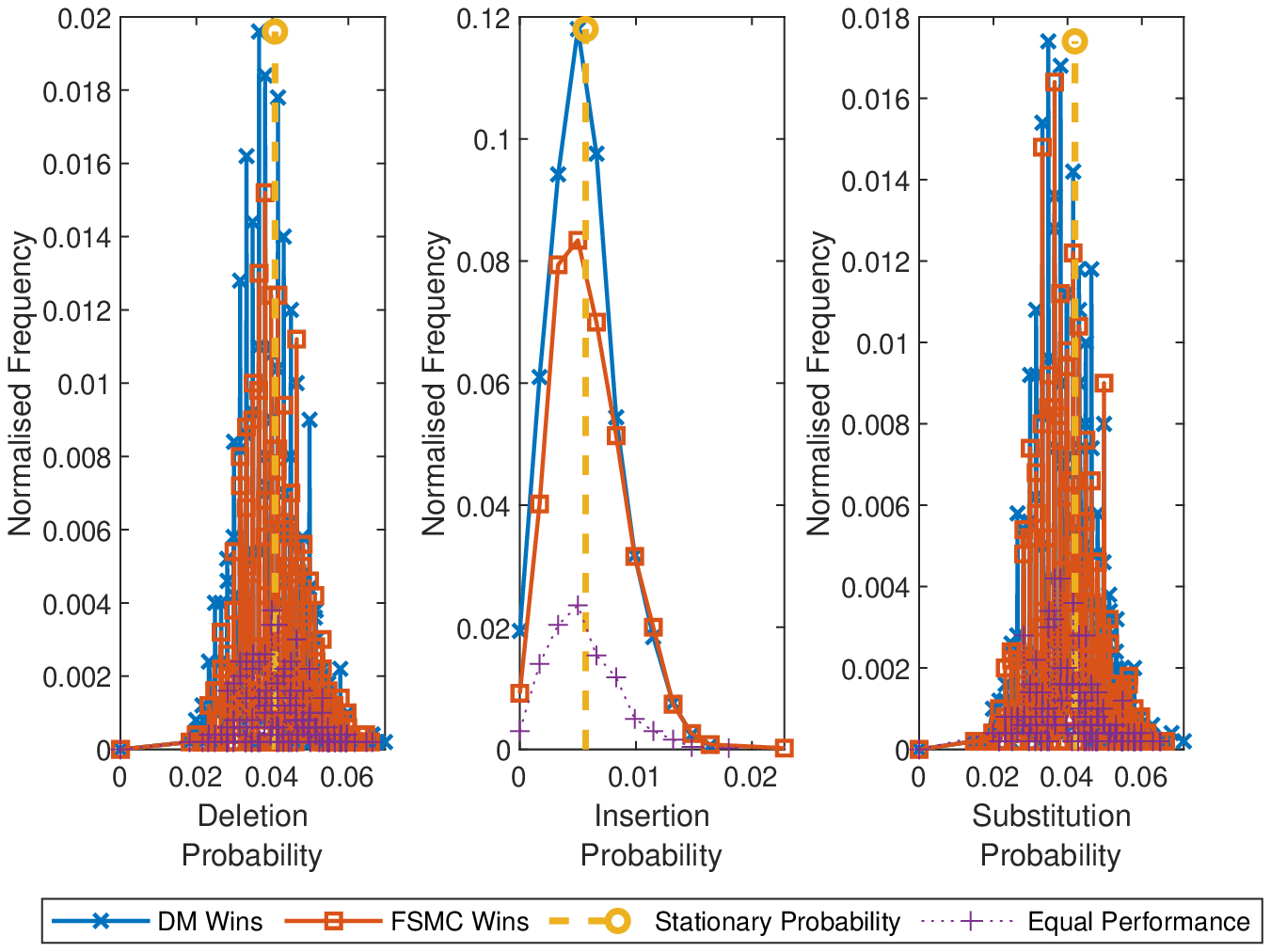}
    \caption[$H=0.292$]{$H=0.292$}   
    \label{fig:freq4}
    \end{subfigure}
    \caption[Frequency decoders outperform each other at various channel error probabilities for a certain entropy value]
    {Frequency decoders outperform each other at various channel error probabilities} 
    \label{fig:freqAnalysis}
    
\end{figure*}

We can further our analysis of the decoding algorithms by taking a closer look at the parameters that contribute to the entropy values and thus affect decoder performance. For the following results, we use the constant entropy data to keep the relevant stationary distributions for a given transition matrix constant. Figure \ref{fig:freqAnalysis} shows the stationary distribution values in yellow against the actual respective frequencies of $P_d$ , $P_i$ and $P_s$ for given entropies. Note that the stationary distribution error probability is scaled to match the maximum frequency of the DM or FSCM plots but has no actual frequency per se. We limit the analysis to certain selected entropy values ($H=0.014$, $0.074$, $0.182$, $0.292$) to illustrate the points. This analysis, however, can be extended to any of the entropy values discussed. From these results, we identify general themes and trends on how the $P_d$, $P_i$ and $P_s$ values affect the decoder performance but note that this is a high-level analysis, and further investigation is required to better identify the prevailing effects of the parameters. In each of the subplots shown in Figure \ref{fig:freqAnalysis}, graphs of the number of times the first-order DM algorithm performs better (blue), the FSCM decoding performs better (red), and the times both the first-order DM and FSCM decoding have identical decoding performance (purple) are plotted. This analysis is based in terms of the NIIS performance metric. From Figure \ref{fig:freq1}, it can be seen that for the majority of cases in the simulated channel use, no errors - be it deletion, insertion or substitution - are witnessed, and it is shown that both the DM and FSCM perform equally for the majority of the simulation runs. There are cases where the DM decoder outperforms the FSCM as it is better equipped to handle no error situations. This is because the FSCM model accounts for memory in the channel and almost always allows for some transition to an error state and consequently has higher probabilities of detecting an error even if none are present. It is worth noting that this does not imply the FSCM is a bad decoder but rather suggests its usefulness is better suited to cases where errors are, in fact, present. In contrast, the DM decoder should be used if the channel is relatively stable and causes little to no synchronisation errors. It can also be argued that a backwards error correction technique such as automatic repeat request would be better suited in these low-error cases than trying to correct these minimal errors. Figure \ref{fig:freq1} shows the outperformance frequency plots at an entropy of $H=0.074$. It is evident that most error probabilities produced by the channel are scattered around the relevant stationary distribution values. When the respective channel error is less than the stationary distribution, it is observed that the DM decoder tends to outperform the FSCM decoder. However, as the probability of the error in question surpasses its corresponding stationary distribution, we notice that the FSCM starts to have a larger number of occurrences where it performs better than the DM decoder. Similar trends are noticed at entropies of $0.182$ and $0.292$, especially concerning the insertion probability graphs. The plots for the entropies of $0.182$ and $0.292$ are shown in Figure \ref{fig:freq3} and \ref{fig:freq4} respectively. It is also noticed that at these higher entropy values, the probability of deletion and substitution caused by the channel spans a larger range of values. In contrast, the range of the insertion probabilities remains relatively constant. Also noted at the higher entropy values is that the decoders are very seldom in agreement with one another, and one decoding algorithm generally outperforms the other at these points. It is still observed that a fair majority of the error probabilities produced by the channel are generally centred around the relevant stationary distribution error probability. For these results and analysis, the second-order DM decoding has been omitted as it closely follows the performance of the first-order decoding for the cases in question.

  %%%%%%%%%%%%%%%%%%%%%%%%%%%%%%%%%%%%%%%%%%%%%%%%%%%%%%%%%  

\subsection{Effects of $P_s$ on simulations}
\begin{figure}[h]
\centering
\includegraphics[width=0.9\columnwidth]{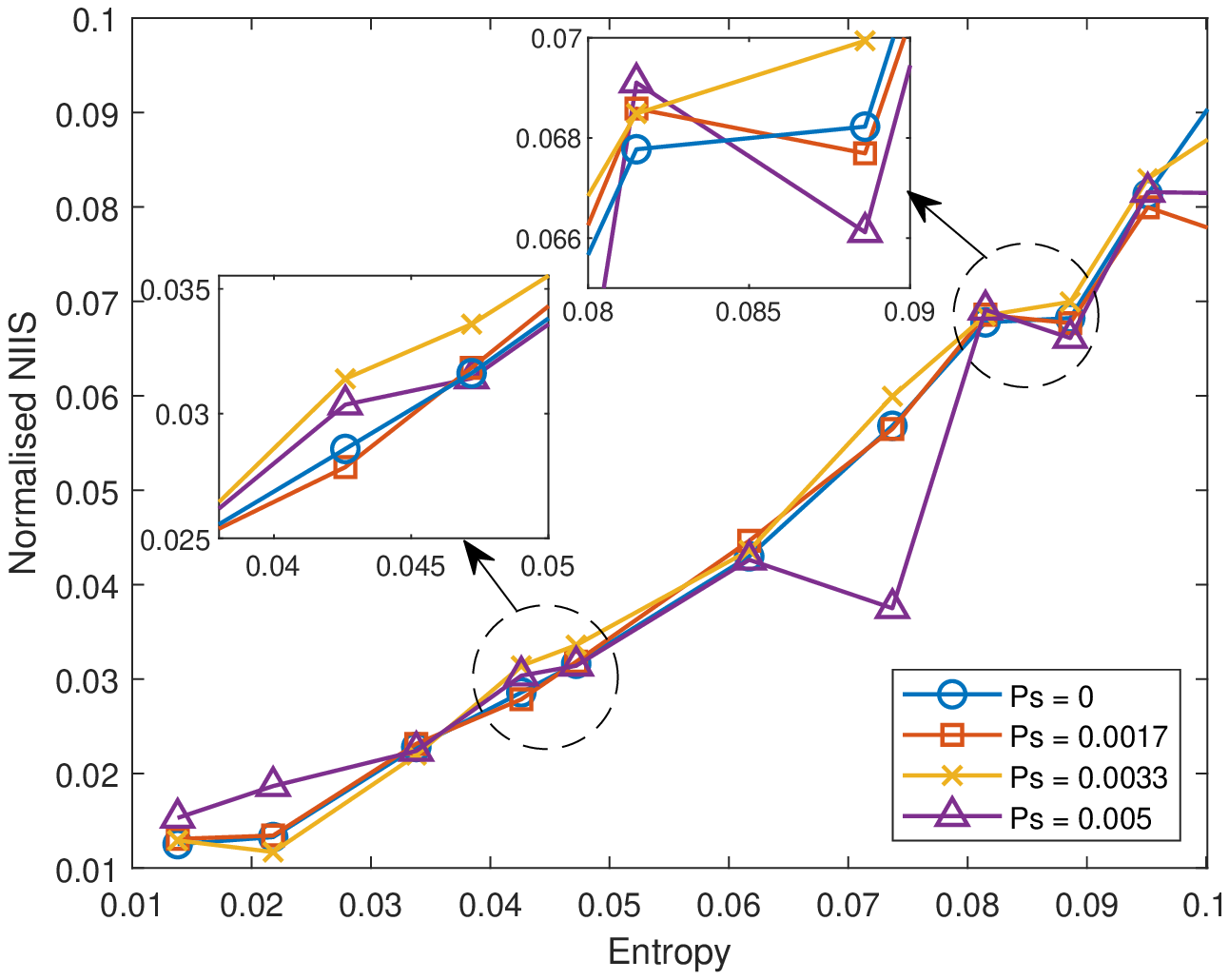}
\caption[Effect of $P_s$ on NIIS]{Effect of $P_s$ on NIIS}
\label{fig:pseffectNIIS}
\end{figure}

As outlined in the proposed system, the entropy values calculated are based on the converted 3-state FSMC and not the original 4-state. In other words, the substitution state is omitted from this calculation as it is converted to a memoryless error value for the discussed system. The following test provides a way to determine the effect the value of $P_s$ has on the system. In these tests, the same transition matrix is used for a given entropy value; however, the value of $P_s$ is varied, and the NIIS values are recorded. The tests show results in an entropy range from $0.01$ to $0.1$ using $P_s$ values of $0.0017$, $0.0033$, $0.005$ and $0.0067$ which corresponds to $1,2,3$ or $4$ substitution errors in the given system. It is worth noting that this is a general excerpt of the results, and higher entropy values will tend to see higher $P_s$ values. Additionally, for given parameters, a certain $P_s$ value may occur more often than others in actual channel usage, as is shown in Figure \ref{fig:freqAnalysis}. As such, the results for this set of tests may contain slight biases. However, these values are normalised according to the number of times a certain $P_s$ value occurred as this tries to minimise the bias. Figure \ref{fig:pseffectNIIS} shows that the value of $P_s$ has a rather random effect on the NIIS, with the system performing better for different cases of $P_s$ and entropy values. However, the normalised NIIS will generally be in a similar region with slight differences for varying $P_s$ values at a set entropy. There are, of course, outliers which are evident at an entropy of $0.0737$ where the NIIS associated with the higher $P_s$ value of $0.005$ has a better performance. However, this is due to the aforementioned bias in the number of occurrences of that specific $P_s$ value. In this case there are 10 counts of $P_s = 0.005$, 3693 counts of $P_s = 0$, 1135 counts of $P_s = 0.0017$ and 162 counts of $P_s = 0.0033$.   

\section{Conclusion}
\label{sec:Conclusion}
The DM synchronisation coding and decoding scheme are discussed along with the FSMC synchronisation channel, which can model an IDS channel while incorporating memory aspects of the channel.  By integrating ideas from both schemes and channels, a novel memory synchronisation system is developed and introduced, which is capable of correcting insertion and deletion errors. Various tests and simulations are done using a variety of evaluation metrics, and it is shown that the new proposed decoder performs similar to that of the original first-order and second-order DM decoder. It is even shown that for specific entropy values (0.01 to 0.1), the proposed decoder performs slightly better than the DM counterparts. A further analysis that delves into the individual error probabilities and their effect on decoder performance is conducted, which reveals a generally better performance from the FSMC proposed decoding when error values exceed that of the corresponding stationary distribution error values. While the coding gain may be insubstantial, the proposed channel is more indicative of real-world communication channels, especially in harsher environments, as errors are more likely to be correlated. Thus the code construction and decoding will likely prove useful for such applications.

\section*{CRediT authorship contribution statement}
\label{sec:CRediT}
\textbf{Shamin Achari}: Conceptualisation, Methodology, Software, Formal analysis, Investigation and Writing - Original Draft. \textbf{Ling Cheng}: Supervision, Conceptualisation and Validation.

\section*{Declaration of Competing Interest}
The authors declare that they have no known competing financial interests or personal relationships that could have
appeared to influence the work reported in this paper.

\section*{Acknowledgements}
This work is based on the research supported in part by the National Research Foundation of South Africa (Grant Numbers: 148765, 132651 \& 129311).

\appendix

\setcounter{table}{0} 
\setcounter{figure}{0} 
\setcounter{equation}{0} 

\section{Generation of Transition Matrices}
\label{app:transmatcreation}
This appendix outlines the procedure to generate various transition matrices which correspond to different ranges of entropy values. The stationary distributions of the four-state matrix, as well as the actual transition matrix entries,  affect the channel entropy. The matrix created is the four-state version which is then converted to a corresponding three-state matrix using the procedure outlined in the main document. In order to create the four-state matrix, different transition probabilities are set up. This is broken into two large groups, firstly a range of probabilities corresponding to a transition from a Transmission to an Error where an error constitutes either substitution, insertion or deletion states. The second range of probabilities is that of Error to Error. Table \ref{tbl:transprob} shows the different error ranges used to generate the matrix entries and the related range of entropy values it produces. Each entry in the matrix is randomly generated within these predefined error ranges and finally, for each row, the entry corresponding to a final Transmission state is calculated by taking 1 minus the other entries in the row. This is done to ensure all possible transitions in a given row equal unity. Once the matrix is created, its equivalent 3-state form is determined and the entropy of the 3-state  model is calculated and used in accordance with the various simulations and tests. As is evident from the values in Table \ref{tbl:transprob}, the higher the probability of staying in a transmission state (lower Transmission to Error and lower Error to Error probability), the lower the entropy. This again intuitively makes sense as the lower the entropy, the less uncertainty is found in the channel. As the entries in the transition matrix become more alike to each other, the entropy increases and it becomes more difficult to determine the correct state as the uncertainty of the channel state is much higher.

\begin{table}[hb]
\centering
\caption{Entropy Ranges and Corresponding Transition Matrix Entries}
\label{tbl:transprob}
%\resizebox{\linewidth}{!}{%
\begin{tabular}{ccc} 
\toprule
Entropy Range & \begin{tabular}[c]{@{}c@{}}Transmission\\to~Error\\Probabilty\end{tabular} & \begin{tabular}[c]{@{}c@{}}Error\\to Error\\Probabilty\end{tabular}  \\
\midrule
0.01 - 0.1    & 0.0001 - 0.005                                                             & 0.001 - 0.05                                                         \\
0.1 - 0.2     & 0.001 - 0.05                                                               & 0.01 - 0.05                                                          \\
0.2 - 0.3     & 0.01 - 0.05                                                                & 0.001 - 0.05                                                         \\
\bottomrule
\end{tabular}
%}
\end{table}

\setcounter{table}{0} 
\setcounter{figure}{0} 
\setcounter{equation}{0} 

\section{Pseudocode for Decoders Forward-Backward Algorithms}
\label{app:algorithms}

This appendix outlines the various forward-backward algorithms used in the testing of the decoders. This allows us to easily illustrate and compare the similarities, and consequently, differences, between the various algorithms used. Algorithm \ref{alg:FBDM1} describes the forward-backward process for the bit level first-order Davey-MacKay decoder. Algorithm \ref{alg:FBFSMC} shows the pseudo-code the forward-backward function for the finite state Markov channel decoder. Lastly, Algorithm \ref{alg:FBDM2} illustrates the forward-backward process for the bit level second-order Davey-MacKay decoder.

\vfill

\begin{algorithm}
\caption{: First-Order DM Forward-Backward Algorithm}
\label{alg:FBDM1}
\begin{algorithmic}[1]
\Function{ForwardBackwardDM}{\textit{maximum offset} $x_{max}$, \textit{maximum consecutive insertions} I}

%\State create Forward $F$, Backward $B$ and ForwardBackward $FB$ probability matrices: \Comment{Setup}
$nStates = [-x_{max},...,x_{max}]$
\State $F[nStates,\Gamma]$,
\State $B[nStates,\Gamma]$,
\State $FB[nStates,\Gamma]$

\ForAll{states $s \in nStates$} 
    \State $F[s,1] \gets \pi[s]$ \Comment{Forward Initialisation}
\EndFor

\ForAll{time $\tau \in [2,...,\Gamma]$} \Comment{Forward Recursion}
    \For{each current state $j \in nStates$} 
        \For{each previous state $i$ from $j-I$ to $j+1$} 
            \If{$i \in nStates$}
                \State $F[j,\tau]\gets \sum F[i,\tau-1](\alpha_{ij} + \beta_{ij}\zeta_{j}^{\tau-1})$
            \EndIf
        \EndFor
    \EndFor
    \State Normalise $F$ across all rows at current $\tau$
\EndFor
%%%%%%%%%%%%%%%%%%%%%%%%%%%%%
\State $B[\rho,\Gamma] \gets 1$ \Comment{Backward Initialisation}

\ForAll {time $\tau \in [\Gamma-1,...,1]$} \Comment{Back Recursion}
    \ForAll{current states $j \in nStstates$} 
        \For{each next state $i \in [j-1,j+I]$} 
            \If{$i \in nStates$}
                \State $B[j,\tau]\gets \sum B[i,\tau+1](\alpha_{ji} + \beta_{ji}\zeta_{j}^{\tau})$
            \EndIf
        \EndFor
    \EndFor
    \State Normalise $B$ across all rows at current $\tau$
\EndFor

\ForAll {$s,\tau \in nStates$} \Comment{FB Calculations}
    \State $FB[s,\tau]\gets F[s,\tau]*B[s,\tau]$
\EndFor
\State Normalise $FB$ across all rows at given $\tau$
\State \Return{FB}
\EndFunction
\end{algorithmic}
\end{algorithm}
%%%%%%%%%%%%%%%%%%%%%%%%%%%%%%%%%%%%%%%%%%%%%%%%%%%%%%%%%%%%%%%%%%%%%%%%

%%%%%%%%%%%%%%%%%%%%%%%%%%%%%%%%%%%%%%%%%%%%%%%%%%%%%%%%%%%%%%%%%%%%%%%%%%%%%%%

\begin{algorithm*}
\caption{: FSMC Forward-Backward Algorithm}
\label{alg:FBFSMC}
\begin{multicols}{2}
  \footnotesize
\begin{algorithmic}[1]
\Function{ForwardBackwardFSMC}{\textit{maximum offset} $x_{max}$, \textit{maximum consecutive insertions} I}

%\State create Forward $F$, Backward $B$ and ForwardBackward $FB$ probability matrices: \Comment{Setup}
$nStates = [-x_{max},...,x_{max}]$
\State $F[nStates,\Gamma]$,
\State $B[nStates,\Gamma]$,
\State $FB[nStates,\Gamma]$

\ForAll{states $s \in nStates$} 
    \State $F[s,1] \gets \pi[s]$ \Comment{Forward Initialisation}
\EndFor

    \For{each current state $j \in nStates$} 
        \For{each previous state $i$ from $j-I$ to $j+1$} 
            \If{$i \in nStates$}
                \State $F[j,2]\gets \sum F[i,1](\alpha_{ij} + \beta_{ij}\zeta_{j}^{1})$
            \EndIf
        \EndFor
    \EndFor
    \State Normalise $F$ across all rows at current $\tau=2$

\ForAll{time $\tau \in [3,...,\Gamma]$} \Comment{Forward Recursion}
    \For{each current state $k \in nStates$} 
        \For{each $\tau-1$ state $j$ from $k-I$ to $k+1$} 
            \For{each $\tau-2$ state $i$ from $j-I$ to $j+1$} 
            \If{$i \And j \in nStates$}
                \State $F[k,\tau]\gets \sum F[j,\tau-1](Prob(i,j,k))$ \Comment{Here $Prob()$ is the relevant probabilities from Table 2}
            \EndIf
            \EndFor
        \EndFor
    \EndFor
    \State Normalise $F$ across all rows at current $\tau$
\EndFor
%%%%%%%%%%%%%%%%%%%%%%%%%%%%%
\State $B[\Psi,\Gamma] \gets 1$ \Comment{Backward Initialisation}

    \ForAll{current states $j \in nStstates$} 
        \For{each next state $i \in [j-1,j+I]$} 
            \If{$i \in nStates$}
                \State $B[j,\Gamma-1]\gets \sum B[i,\Gamma](\alpha_{ji} + \beta_{ji}\zeta_{j}^{\Gamma-1})$
            \EndIf
        \EndFor
    \EndFor
    \State Normalise $B$ across all rows at current $\tau$

\ForAll {time $\tau \in [\Gamma-2,...,1]$} \Comment{Back Recursion}
    \ForAll{current states $k \in nStstates$} 
        \For{each $\tau+1$ state $j \in [k-1,k+I]$} 
            \For{each $\tau+2$ state $i \in [j-1,j+I]$} 
            \If{$i \And j \in nStates$}
                \State $B[k,\tau]\gets \sum B[j,\tau+1](Prob(k,j,i))$ \Comment{Here $Prob()$ is the relevant probabilities from Table 2}
            \EndIf
            \EndFor
        \EndFor
    \EndFor
    \State Normalise $B$ across all rows at current $\tau$
\EndFor

\ForAll {$s,\tau \in nStates$} \Comment{FB Calculations}
    \State $FB[s,\tau]\gets F[s,\tau]*B[s,\tau]$
\EndFor
\State Normalise $FB$ across all rows at given $\tau$
\State \Return{FB}

\EndFunction
\end{algorithmic}
\end{multicols}
\end{algorithm*}

\begin{algorithm*}
\caption{: Second-Order DM Forward-Backward Algorithm}
\label{alg:FBDM2}
\begin{multicols}{2}
\footnotesize
\begin{algorithmic}[1]
\Function{2ndOrderForwardBackwardDM}{\textit{maximum offset} $x_{max}$, \textit{maximum consecutive insertions} I}

%\State create Forward $F$, Backward $B$ and ForwardBackward $FB$ probability matrices: \Comment{Setup}
$nStates = [-x_{max},...,x_{max}]$
\State $F[nStates,\Gamma]$,
\State $B[nStates,\Gamma]$,
\State $FB[nStates,\Gamma]$

\ForAll{states $s \in nStates$} 
    \State $F[s,1] \gets \pi[s]$ \Comment{Forward Initialisation}
\EndFor

\For{each current state $j \in nStates$} 
        \For{each previous state $i$ from $j-I$ to $j+1$} 
            \If{$i \in nStates$}
                \State $F[j,2]\gets \sum F[i,1](\alpha_{ij} + \beta_{ij}\zeta_{j}^{1})$
            \EndIf
        \EndFor
    \EndFor
    \State Normalise $F$ across all rows at current $\tau$

\ForAll{time $\tau \in [3,...,\Gamma]$} \Comment{Forward Recursion}
    \For{each current state $k \in nStates$} 
        \For{each $\tau-1$ state $j$ from $k-I$ to $k+1$} 
        \For{each $\tau-2$ state $i$ from $j-I$ to $j+1$}
            \If{$i \And j \in nStates$}
                \State $F[k,\tau]\gets \sum F[j,\tau-1]((\alpha_{ij} + \beta_{ij}) * \alpha_{jk} + \beta_{jk}\zeta_{k}^{\tau-1})$
            \EndIf
            \EndFor
        \EndFor
    \EndFor
    \State Normalise $F$ across all rows at current $\tau$
\EndFor
%%%%%%%%%%%%%%%%%%%%%%%%%%%%%

\State $B[\Psi,\Gamma] \gets 1$ \Comment{Backward Initialisation}

\For{each current state $j \in nStates$} 
        \For{each next state $i$ from $j-1$ to $j+I$} 
            \If{$i \in nStates$}
                \State $B[j,\Gamma-1]\gets \sum [j,\Gamma](\alpha_{ji} + \beta_{ji}\zeta_{j}^{\Gamma-1})$
            \EndIf
        \EndFor
    \EndFor
    \State Normalise $B$ across all rows at current $\tau$

\ForAll {time $\tau \in [\Gamma-2,...,1]$} \Comment{Back Recursion}
    \ForAll{current states $k \in nStstates$} 
        \For{each $\tau+1$ state $j \in [k-1,k+I]$} 
            \For{each $\tau+2$ state $i \in [j-1,j+I]$}
            \If{$i \And j \in nStates$}
                \State $B[k,\tau]\gets \sum B[j,\tau+1]((\alpha_{ji} + \beta_{ji}) * \alpha_{kj} + \beta_{kj}\zeta_{k}^{\tau})$ 
            \EndIf
            \EndFor
        \EndFor
    \EndFor
    \State Normalise $B$ across all rows at current $\tau$
\EndFor

\ForAll {$s,\tau \in nStates$} \Comment{FB Calculations}
\State $FB[s,\tau]\gets F[s,\tau]*B[s,\tau]$
\EndFor
    \State Normalise $FB$ across all rows at given $\tau$
\State \Return{FB}
\EndFunction

\end{algorithmic}
\end{multicols}
\end{algorithm*}
%%%%%%%%%%%%%%%%%%%%%%%%%%%%%%%%%%%%%%%%%%%%%%%%%%%%%%%%%%%%%%%%%%%%%%%%

 \bibliographystyle{elsarticle-num} 
 \bibliography{ref}





% \end{thebibliography}

\end{document}